\documentclass[12pt]{iopart}

\usepackage{multicol}
\usepackage{graphicx}
\usepackage{booktabs}
\usepackage{amssymb,bm,mathrsfs,bbm,amscd}
\usepackage{lastpage}
\usepackage{CJK}

\def\bd{\begin{document}} \def\ed{\end{document}}
\def\bmp{\begin{minipage}} \def\emp{\end{minipage}}
\def\bcc{\begin{center}} \def\ecc{\end{center}}     \def\npg{\newpage}
\def\beq{\begin{equation}} \def\eeq{\end{equation}} \def\hph{\hphantom}
 \def\r#1{$^{[#1]}$}
\def\n{\noindent} \def\ni{\noindent} \def\pa{\parindent}
\def\hs{\hskip} \def\vs{\vskip} \def\hf{\hfill} \def\ej{\vfill\eject}
\def\cl{\centerline} \def\ob{\obeylines}  \def\ls{\leftskip}
\def\underbar#1{$\setbox0=\hbox{#1} \dp0=1.5pt \mathsurround=0pt
   \underline{\box0}$}   \def\ub{\underbar}    \def\ul{\underline}
\def\f{\left} \def\g{\right} \def\e{{\rm e}} \def\o{\over} \def\d{{\rm d}}
\def\vf{\varphi} \def\pl{\partial} \def\cov{{\rm cov}} \def\ch{{\rm ch}}
\def\la{\langle} \def\ra{\rangle} \def\EE{e$^+$e$^-$} \def\pt{p_{\rm T}}
\def\pti{p_{{\rm T},i}} \def\yti{y_{{\rm T},i}}
\def\ptj{p_{{\rm T},j}}\def\mt{m_{\rm T}} \def\yt{y_{\rm T}} \def\vt{v_{\rm T}}

\def\bitz{\begin{itemize}} \def\eitz{\end{itemize}}
\def\btbl{\begin{tabular}} \def\etbl{\end{tabular}}
\def\btbb{\begin{tabbing}} \def\etbb{\end{tabbing}}
\def\beqar{\begin{eqnarray}} \def\eeqar{\end{eqnarray}}
\def\\{\hfill\break} \def\dit{\item{-}} \def\i{\item}
\def\bbb{\begin{thebibliography}{9} \itemsep=-1mm}
\def\ebb{\end{thebibliography}} \def\bb{\bibitem}
\def\bpic{\begin{picture}(260,240)} \def\epic{\end{picture}}
\def\akgt{\cl{\bf ACKNOWLEDGMENTS}}
\def\fgn{\noindent{\bf\large\bf figure captions}}
\def\lan{\langle}
\def\ran{\rangle}
\def\p{\pi}
\def\ifmath#1{\relax\ifmmode #1\else $#1$\fi}%
\def\rc{\ifmath{{\mathrm{c}}}}
\def\cut{\ifmath{{\mathrm{cut}}}}
\def\rF{\ifmath{{\mathrm{F}}}}
\def\rK{\ifmath{{\mathrm{K}}}}
\def\rp{\ifmath{{\mathrm{p}}}}
\def\rt{\ifmath{{\mathrm{t}}}}
\def\LAB{\ifmath{{\mathrm{LAB}}}}
\def\cut{\ifmath{{\mathrm{cut}}}}

\newcommand{\gguide}{{\it Preparing graphics for IOP journals}}
\begin{document}
%\begin{CJK*}{GBK}{song}

\title[]{Several problems on the measured  hyperorder cumulants of net-proton distributions  in heavy-ion collisions}

%\author{Lizhu Chen$^{1*}$, Yuhan Wang$^1$, Ye -Yin Zhao$^2$, Zhiming Li$^3$, and Yuanfang Wu$^{3}$}
\author{Lizhu Chen$^{1*}$, Ye -Yin Zhao$^2$, Zhiming Li$^3$, and Yuanfang Wu$^{3}$}
\address{$^1$ School of Physics and Optoelectronic Engineering, Nanjing University of Information Science and Technology, Nanjing 210044, China}
\address{$^2$ School of Physics and Electronic Engineering, Sichuan University of Science and Engineering (SUSE), Zigong 643000, China}
\address{$^3$ Key Laboratory of Quark and Lepton Physics (MOE) and Institute of Particle Physics, Central China Normal University, Wuhan 430079, China}
\address{$*$ Email: chenlz@nuist.edu.cn}

\begin{abstract}
Hyperorder cumulants $C_5/C_1$ and $C_6/C_2$ are recommended as sensitive observables
 to  explore the QCD phase transition in heavy-ion collisions. 
 Precisely measuring their results remains a difficult task in experiments, when employing
 the  Centrality Bin Width Correction (CBWC) to suppress the initial volume fluctuations.
 Various techniques within the CBWC formula
 can lead to notable differences in the results. 
We will systematically  investigate the application of the CBWC method to the measured 
net-proton $C_5/C_1$ and $C_6/C_2$  using the UrQMD model and Skellam-based simulations at  $\sqrt{s_{NN}}$ = 11.5 GeV in Au + Au collisions.  
A recommended approach is proposed to calculate $C_5/C_1$ and $C_6/C_2$ in 0-40\% centrality.
With statistics comparable to the RHIC Beam Energy Scan phase II (BES-II), 
our studies  provide a baseline for analyzing net-proton $C_5/C_1$ and $C_6/C_2$ in relativistic heavy-ion collisions.

\end{abstract}


\section{Introduction}
\vspace{1mm}

High-order cumulants of conserved quantities, such as net-baryon, net-charge and net-strangeness,  are sensitive observables for investigating the QCD 
phase diagram~\cite{QCD, cumulant-1, cumulant-2, cumulant-3, cumulant-4, cumulant-5}.
Lattice QCD calculations show that the fifth-to-first order ($C_5/C_1$) and the sixth-to-second order ($C_6/C_2$) order cumulant ratios exhibit a negative signal 
in the crossover region~\cite{C6-karsch-v1, C6-karsch-v2, C6-karsch-v3, C6-karsch-v4}.
The QCD-assisted low-energy effective theory and the QCD based models, such as the Polyakov loop extended quark-meson (PQM) and the Nambu-Jona-Lasinio (PNJL) models, also support
the idea that  these two ratios turn
 negative near the chiral crossover transition~\cite{PQM-C6-v0,PQM-C6-v1,PQM-C6-v2, PNJL-C6-v1, PNJL-C6-v2}. 

In experiment, the STAR Collaboration has reported the beam energy and collision centrality dependence of net-proton $C_5/C_1$ and $C_6/C_2$ in Au + Au collisions at center-of-mass energies $\left(\sqrt{s_{NN}}\right)$
from 3 to 200 GeV~\cite{STAR-C6C2-v1, STAR-C6C2-v2}.  
Except at 3 GeV, the measured $C_6/C_2$
values for 0-40\% centrality collisions display a progressively
negative trend with decreasing energy. 
 Conversely, $C_5/C_1$ for
0-40\% centrality collisions from 7.7 to 200 GeV exhibits a
weak beam energy dependence and fluctuates about zero.
However, these behaviors can also be theoretically explained without taking into account the QCD phase transition.
The negative  $C_6/C_2$ is also noted in the sub-ensemble acceptance method~\cite{sam-1, sam-2, sam-3} and hydrodynamics calculations~\cite{hydrodynamics}, encompassing
global conservation, system non-uniformity, and momentum space acceptance.
In addition, both the experimental behaviors of these two ratios can also be attributed to low statistics effects without necessitating the consideration of any phase transition~\cite{lizhu-1, lizhu-2}.
 Consequently, It is important to accurately  measure these two ratios and eliminate the impacts of non-phase transitions.~\cite{non-c1, non-c2, non-c4, non-c5, non-c6, non-c7}.

  Currently, removing the initial volume fluctuations remains one of the most challenging issues in accurately measuring the cumulants~\cite{Non-V-1, Non-V-2, Non-V-3}.
Experimentally,  the impact of initial volume fluctuations is typically mitigated using the centrality bin width correction (CBWC) method~\cite{cbwc-v1, cbwc-v2, cbwc-v3}. 
This is achieved by computing cumulants or cumulant ratios in each multiplicity bins before combining the results within a specific centrality bin width, such as
0-10\%, 10-20\%, 20-30\%,$\cdots$, 70-80\% centralities. The effectiveness of the CBWC method has been a topic of extensive discussion. 
In the case of net-proton $C_6/C_2$ and $C_5/C_1$, the measured results exhibit a strong statistical dependence with 
insufficient statistics, and the CBWC method exacerbates the impact of low statistics on the measured hyperorder cumulant~\cite{chenlz-npa, lizhu-1, lizhu-2}.
In addition, when aggregating results into a specific centrality, 
the event-weighted scheme has proven successful for cumulants up to the sixth order, while the error-weighted scheme may lead to underestimation of the measurement~\cite{cbwc-v4, cbwc-v5, koch-error}.

Nevertheless, to precisely quantify the hyperorder cumulants, there are still various specific intricacies of the CBWC method that necessitate in-depth discussions.
In particular, a wider centrality bin width of  0-40\% is constructed for analyzing $C_5/C_1$ and  $C_6/C_2$ to reduce uncertainties. 
This adjustment could enhance the sensitivity of the measurements to the specific schemes of the CBWC method.
When utilizing the CBWC method to calculate $C_6/C_2$ (and also $C_5/C_1$),  one has the option to  either compute $C_6$ and $C_2$ separately or to directly calculate $C_6/C_2$ in each multiplicity bin.
In this paper, we will demonstrate that choosing the latter approach results in the volume cancellation effect being more closely aligned with theoretical predictions, 
and the related statistical uncertainties are significantly reduced.
This reduction in uncertainties is particularly beneficial  for analyzing $C_6/C_2$ and $C_5/C_1$ in experiment, even with the collected statistics at RHIC/BES-II.

Moreover, after calculating  $C_6/C_2$ and $C_5/C_1$ in 0-10\%, 10-20\%, 20-30\% and 30-40\% centralities,
the consequences of constructing these two ratios in 0-40\% centrality
by using error-weighted or event-weighted schemes will also require examination.
It will be demonstrated that the event-weighted approach is better suited for analyzing $C_5/C_1$ and $C_6/C_2$ using minibias data samples at RHIC/BES-II.
Implementing the error-weighted scheme leads to that the result in  0-10\% 
has no any contribution to the  final value in 0-40\% centrality. 
The results in 10-40\% and 0-40\% centralities are identical, encompassing both the measured central value and its associated statistical uncertainty.
This lack of contribution from the 0-10\% centrality 
is mainly attributed to the larger uncertainty  in this centrality compared to the others.

The paper is organized as follows. In Section II, we will offer a thorough explanation and provide formulas for the different CBWC methods.
The results derived from the UrQDM model and Skellam-based Monte Carlo simulations will be
 demonstrated in section III. Finally, a summary is given in section IV.

\section{Various CBWC methods}

As we know, in thermal equilibrium, cumulants of the conserved charge are connected to the thermodynamic susceptibilities
 which are determined by taking derivatives of the logarithm of the QCD partition functions with respect to the chemical potentials of conserved charges:
\begin{equation}\label{cumulants-suscep}
C_{n}^{X}  = \left(VT^3\right) \cdot \chi_n = \frac{\partial ^n \ln Z}{\partial \left (\mu_X/T \right)^n}, \; \; \; \; \;\;\;X = B, Q, S. 
\end{equation}
The formula of the cumulants are expressed as follows:
\begin{eqnarray}\label{Cx-definition}
C_1 &=& \langle N \rangle, \\
C_2 &=& \langle (\delta N)^2 \rangle,  \\
C_3 &=& \langle (\delta N)^3 \rangle,  \\
C_4 &=& \langle (\delta N)^4 \rangle -3\langle (\delta N)^2 \rangle^2, \\
C_5 &=& \langle (\delta N)^5 \rangle - 10  \langle (\delta N)^2 \rangle \langle (\delta N)^3 \rangle, \label{eq:5}\\
%\end {eqnarray}
%\begin{eqnarray}
C_6 &=& \langle (\delta N)^6 \rangle + 30 \langle (\delta N)^2 \rangle^3 \nonumber \\
& &-15\langle (\delta N)^2 \rangle \langle (\delta N)^4 \rangle -10 \langle (\delta N)^3 \rangle^2, \label{eq:6}
\end{eqnarray}
here $\delta N = N - \langle N \rangle$, $N$ is the number of particles in one event, and the average is taken over all events.
Eq.~(\ref{cumulants-suscep}) demonstrates that cumulants are extensive variables that are directly proportional to the volume of the collision system and thermal susceptibilities. 
Ignoring the impact of temperature fluctuation effects, the cumulant ratios, 
given by
 \begin{equation}\label{cumulants-ratio}
\frac{C_m}{C_n}  = \frac{\chi_m}{\chi_n}
\end{equation}
are utilized to study of the QCD phase transition, as they are sensitive to it and can also mitigate the volume contributions.
A Skellam distribution, defined as the difference between two independent Poisson distributions, is typically utilized as a statistical baseline~\cite{Skellam-1, Skellam-2, Skellam-3}, which is unity for $C_4/C_2$, $C_5/C_1$, and $C_6/C_2$.

Experimentally, due to the impact of initial volume fluctuations, 
the cumulants are not directly  calculated in a given centrality bin, such as  0-5\%, 5-10\% and 30-40\% centralities.
The CBWC method is applied to suppress the the initial volume fluctuations on the measured cumulants ~\cite{cbwc-v1,cbwc-v2,cbwc-v3}.
 The procedure is mathematically expressed in the equation below: 
\begin{equation}\label{CBWC-1-1}
C_{n} = \frac{\sum_r n_r C_n^{r}}{\sum_r n_r} = \sum_r \omega_rC_{n}^{r}
\end{equation}
where $n_r$ is the number of events in the $r$th multiplicity bin for the centrality determination.
The corresponding weight  for the $r$th 
multiplicity bin is $\omega_r = n_r/\sum_r n_r$.
$C_n^{r}$ represents the $n-$th order cumulant of particle number distributions in the $r$th multiplicity bin. 
In this case, the cumulant ratio in a finite centrality interval can be described as :
\begin{equation}\label{Cmn}
\frac{C_m}{C_n}= \frac{\sum_r n_r C_{m}^{r}}{\sum_r n_r C_{n}^{r}} = \frac{\sum_r n_r V_r \chi_{m}^{r}}{\sum_r n_r V_r \chi_{n}^{r}} 
\end{equation}
This method is denoted as  the CBWC-I method in this paper.
However, the impact of volume elimination
shown in Eq.~(\ref{Cmn}) does not entirely align with the theoretical expectations as illustrated in  Eq.~(\ref{cumulants-ratio}).

Alternatively, we can directly calculate $\left(C_m/C_n\right)^{r}$ and then obtain the centrality dependency using the corresponding formula is : 
\begin{eqnarray}\label{CBWC-1-2}
C_m/C_n & =& \frac{\sum_r n_r \left(C_m/C_n\right)^{r}}{\sum_r n_r} = \sum_r \omega_r\left(C_m/C_n\right)^{r} \nonumber\\
                & =& \sum_r \omega_r\left(\chi_m/\chi_n\right)^{r} 
\end{eqnarray}
This is referred to as CBWC-II method in this paper. 
It is clear that this formula aligns more closely with the theoretical expectations outlined in Eq.~(\ref{cumulants-ratio}).

 Essentially, both the CBWC-I and -II methods correspond to the event-weighted average.
 Experimentally, the error-weighted average is also employed for data analysis in experiments, which can result in smaller statistical uncertainties.
 Nonetheless, when constructing the results in each multiplicity bin to eight centralities, 
 several studies have suggested that the error-weighted average can underestimate the measured cumulant ~\cite{cbwc-v4, cbwc-v5}. 
 In this paper, we will further investigate whether the ratios in the 0-40\% centrality can be averaged by considering the results in 
 the 0-10\%, 10-20\%, 20-30\%, and 30-40\% centralities using the error-weighted scheme.
 the corresponding formula is :
 \begin{equation}\label{error_weight_average}
C_m/C_n  = \frac{\sum_j\frac{1.0} { \left(\sigma_{C_m/C_n}\right)_j^{2}}\left(C_m/C_n\right)_j}{\sum_j \frac{1.0} { \left(\sigma_{C_m/C_n}\right)_j ^{2}}} 
\end{equation}
Here, $(C_m/C_n)_j$ and  $\left(\sigma_{C_m/C_n}\right)_j$  represent the cumulant ratio and its associated error within the specified four centrality bins.
Disregarding the uncertainty of $\left(\sigma_{C_m/C_n}\right)_j$, the  error of $C_m/C_n$ in 0-40\% centrality 
can be determined using the error propagation formula, as follows:
 \begin{equation}\label{error_weight_error}
error  = \frac{1.0}{\sqrt{\sum_j \frac{1.0} { \left(\sigma_{C_m/C_n}\right)_j ^{2}}}} 
\end{equation}
It is called the CBWC-III method in this paper.
 
\section{Comparisons and Discussions}

We analyze how these three CBWC methods affect the net-proton
$C_5/C_1$ and $C_6/C_2$ using the UrQMD model (version 3.4)~\cite{UrQMD-1, UrQMD-2}. 
 We  generate a total of 125M minimum bias events in Au + Au collisions at $\sqrt{s_{NN}}$ = 11.5 GeV.  
 In an effort to avoid auto-correlations and improve the centrality resolutions, 
 this study utilizes the Refmult3X multiplicity, which is calculated based on the count of charged pions and kaons 
  within the pseudorapidity range of $|\eta|<1.6$ in each event. This criterion is specified by the RHIC/BES-II cumulant analysis~\cite{BES-II-cumulant}.
  The kinematic intervals of transverse momentum and rapidity are $0.4<p_{T}<2.0$ GeV/$c$ and $|y|<0.5$. The statistical error is estimated by the Bootstrap method~\cite{Bootstrap-1, Bootstrap-2}.  
 
 \begin{figure}
\centering
\includegraphics[width=3.8in]{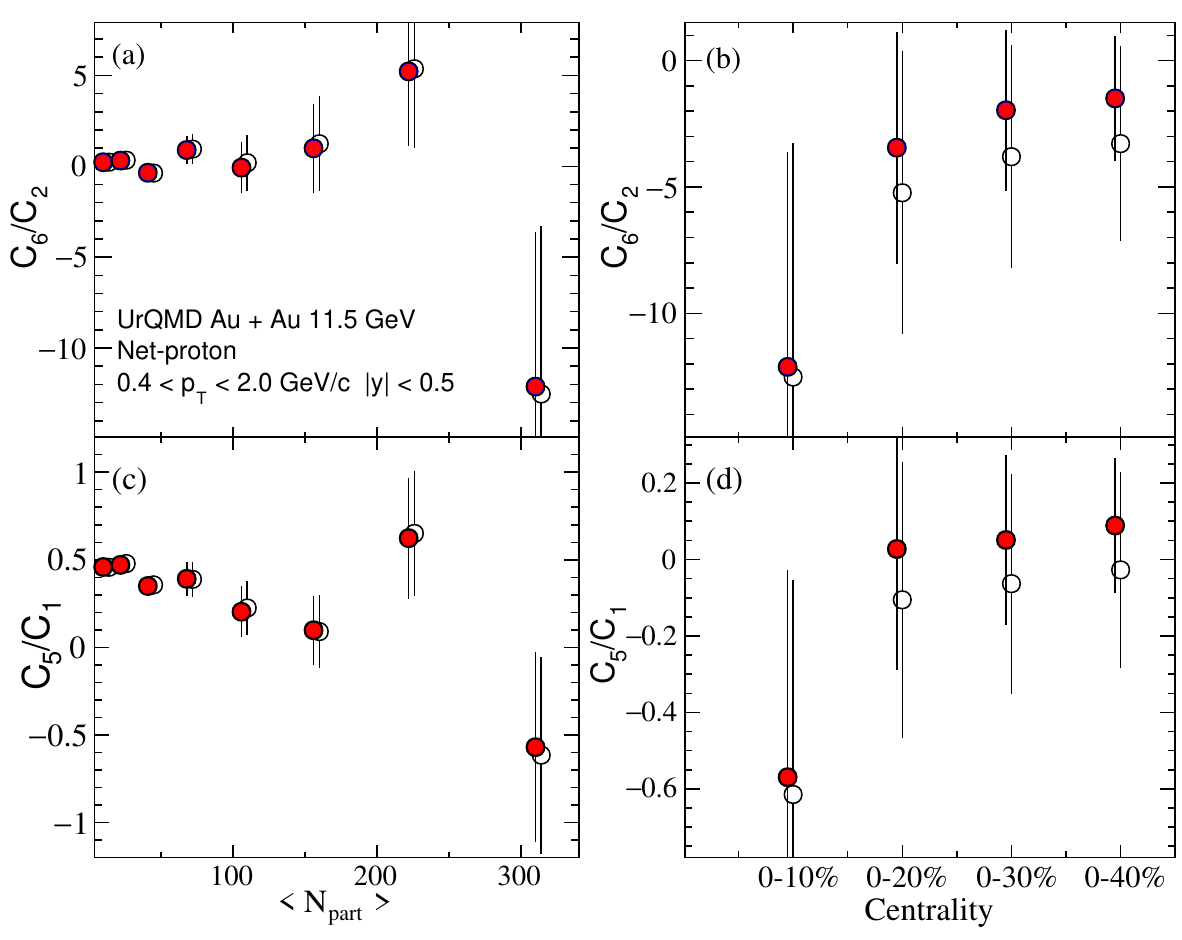}
\caption{\label{c62_cbwc_urqmd} (Color Online)  
The ratios of net-proton $C_6/C_2$ and $C_5/C_1$ in the UrQMD model in eight centralities (left panel) and different centrality bin widths (right panel)  using  CBWC-I (open circles) and CBWC-II (red solid circles) methods. 
  The results for the same centrality are slightly off-set horizontally (including all the following plots) to improve clarity.
}
\end{figure}

\subsection{CBWC-I and -II methods}
Fig.~\ref{c62_cbwc_urqmd}  shows the net-proton ratios $C_6/C_2$ and $C_5/C_1$ using CBWC-I and  CBWC-II methods, 
denoted by $\left(C_6/C_2\right)_{(C-I)}$, 
$\left(C_6/C_2\right)_{(C-II)}$,  $\left(C_5/C_1\right)_{(C-I)}$ and $\left(C_5/C_1\right)_{(C-II)}$, respectively.
The results shown are for  the eight centralities (0-10\%, 10-20\%, 20-30\% $\cdots$ 70-80\%) and different centrality bin widths (0-10\%, 0-20\%, 0-30\%, and 0-40\%).
  In the case of the eight centralities, 
  the discrepancy between the CBWC-I and CBWC-II methods in calculating the ratios
 $C_6/C_2$ and $C_5/C_1$
 is  negligible for all centralities, including both the measured values and the estimated statistical uncertainties.
 However, with the widening of the centrality bin width, variations in the measured values for both
 $C_6/C_2$ and $C_5/C_1$ are observed.
  The discrepancies in the measured uncertainties become more pronounced as the centrality bin width increases.
  In 0-40\% centrality, the errors of 
$\left(C_6/C_2\right)_{(C-II)}$ and $\left(C_5/C_1\right)_{(C-II)}$  are 68.8\% and 63.7\%, respectively,
compared to those of $\left(C_6/C_2\right)_{(C-I)}$ and $\left(C_5/C_1\right)_{(C-I)}$.
Therefore, the impact of the decreased uncertainties for the CWBC-II method is significantly in the UrQMD calculations.

 \begin{figure}
\centering 
\includegraphics[width=4.8in]{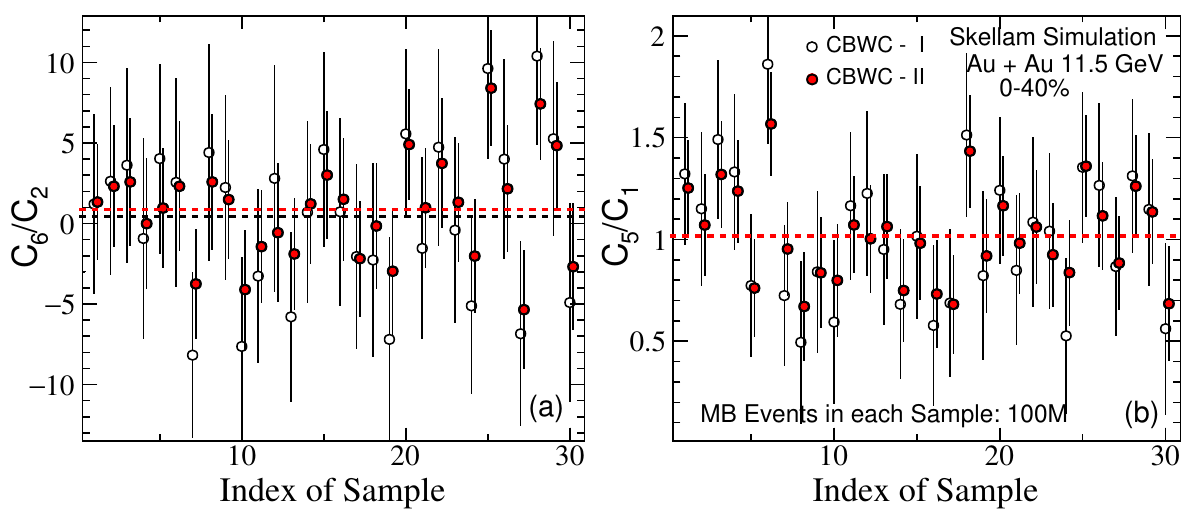}
\includegraphics[width=4.8in]{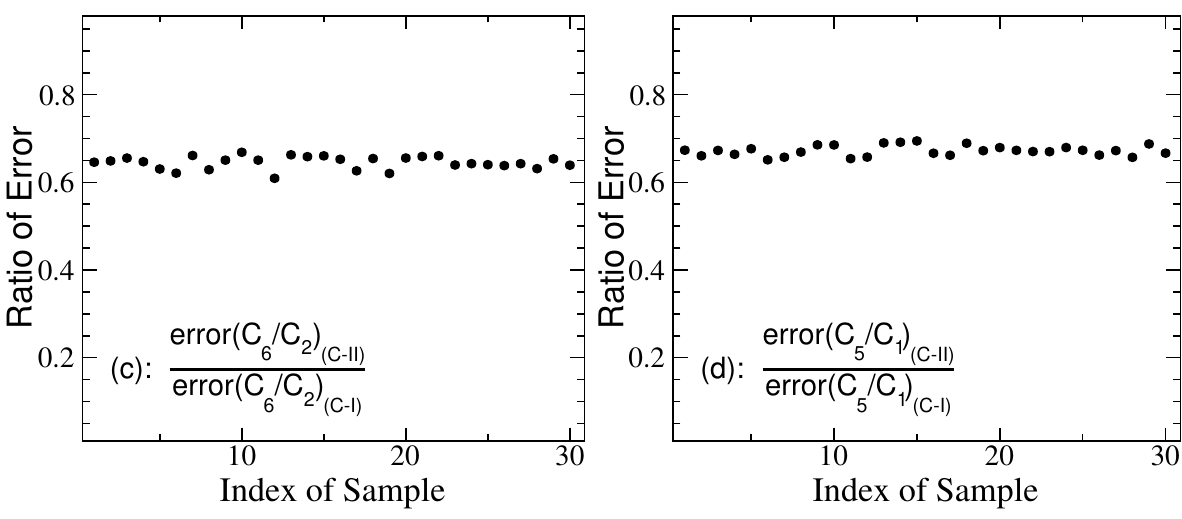}
\caption{\label{c62_c51_simulation} (Color Online)  
Skellam-based simulations of net-proton (a) $C_6/C_2$ and (b) $C_5/C_1$  for $0-40\%$ Au + Au collisions at $\sqrt{s_{NN}}$ = 11.5 GeV, utilizing 100 million MB events in each sample. 
The results shown are for CBWC-I (open circles) and CBWC-II (red solid circles) methods.
The Refmult3X distribution and $\left<N_{p}\right>$ and $\left<N_{\bar{p}}\right>$ for each Refmult3X are taken from the UrQMD model. 
The  averages in 0-40\% centrality  over all 30 samples  by CBWC-I and CBWC-II are represented by the black and red dashed lines, respectively.
The respective error ratios calculated using CBWC-II and CBWC are shown in (c) and (d), respectively.

}
\end{figure}

To comprehensively assess the effectiveness of these two methods at a given statistics and centrality bin, we further conducted extensive Skellam-based Monte Carlo simulations.
30 samples have been generated, each consisting of $100\times 10^{6}$ (100M) minimum bias (MB) events.  
In each event, the simulated Refmult3X follows the same distributions as the UrQMD model, 
and 
 $N_{p}$ and $N_{\bar{p}}$ obey Poisson distributions, with $\left<N_{p}\right>$ and $\left<N_{\bar{p}}\right>$ obtained from the corresponding Refmult3X bin in the UrQMD model.
 Hence, the simulated data we generated are minimum bias samples, and the net-proton number follows the Skellam distributions in each Refmult3X bin.
  Consequently, the excepted values of $C_5/C_1$ and $C_6/C_2$ in 0-40\% centrality should be unity.
  
In Fig.~\ref{c62_c51_simulation},
the simulated $C_6/C_2$ (left panel) and $C_5/C_1$ (right panel) values in 0-40\% centrality by these two methods are denoted by 
 $\left(C_6/C_2\right)_{(C-I)}$, $\left(C_6/C_2\right)_{(C-II)}$, $\left(C_5/C_1\right)_{(C-I)}$, and $\left(C_5/C_1\right)_{(C-II)}$, respectively. 
 The corresponding  averages of  these four variables  in 0-40\% centrality over all  30 samples are represented by black and red dashed lines in left and right panels, respectively.
Indeed, these four  averages are all near unity, aligning with the Skellam baseline. 
Here the averages of $\left(C_5/C_1\right)_{(C-I)}$ and $\left(C_5/C_1\right)_{(C-II)}$ almost identical, resulting in the black and red dashed lines overlapping in the right panel.
 All four variables display random fluctuations around their average values. The measured values and estimated uncertainties follow statistical principles. 
 
 In Fig.~\ref{c62_c51_simulation}(a),  the differences between each $\left(C_6/C_2\right)_{(C-I)}$ and $\left(C_6/C_2\right)_{(C-II)}$ is visible.
 Above the line of average,  the value of each $\left(C_6/C_2\right)_{(C-II)}$
 lie below its corresponding $\left(C_6/C_2\right)_{(C-I)}$. Below the line of the average value, the value of each $\left(C_6/C_2\right)_{(C-II)}$ is higher 
 than that of $\left(C_6/C_2\right)_{(C-I)}$. In particular, below 1$\sigma$ of the average (where the error bar does not touch the expectation line), the value of each $\left(C_6/C_2\right)_{(C-II)}$  is significantly  closer to the 
 theoretical expectation.
 Meanwhile,  Fig.~\ref{c62_c51_simulation}(b) shows similar results for $\left(C_5/C_1\right)_{(C-I)}$ and $\left(C_5/C_1\right)_{(C-II)}$.
   Therefore,  with a given statistics, these two ratios calculated by CBWC-II method is more close to its expectation in 0-40\% centrality.
 
For these two ratios, Fig.~\ref{c62_c51_simulation} also shows that the errors obtained by CBWC-II 
  are significantly smaller than 
those obtained by CBWC-I.
The ratios  of $\frac{error(C_6/C_2)_{(C-II)}}{error(C_6/C_2)_{(C-I)}}$ and $\frac{error(C_5/C_1)_{(C-II)}}{error(C_5/C_1)_{(C-I)}}$ are shown in  Fig.~\ref{c62_c51_simulation}(c) and (d), respectively.  For these two ratios, 
all errors obtained using CBWC-II are approximately  $60\% \sim 70\%$  compared to those obtained using CBWC-I. 
This decrease in statistical error also aligns with what was demonstrated in the UrQMD results shown in Fig.~\ref{c62_cbwc_urqmd}.
As we know, when the statistical sample doubles, the associated statistical uncertainty becomes approximately  $\frac{1}{\sqrt{2}}$($\sim$70.7\%)  of the original.
Therefore, the reduction in uncertainties with the CBWC-II method is akin to doubling the statistical sample in experiment.
 
 In addition, the formula of CBWC-II method is more equivalent to the theoretical formula as shown in Eq.~(\ref{CBWC-1-2}).
 Consequently,  the CBWC-II method is  a more suitable calculation
method for data analysis in 0-40\% centrality.  So in comparison to the CBWC-I, the CBWC-II method more precisely 
measures the hyperorder cumulants.

\subsection{Influence of weighted factor in average}  

Before implementing the CBWC-III method, let's first discuss  why the error-weighted average can underestimate the measured 
 values of $C_6/C_2$ and $C_5/C_1$ in the  0-10\%, and $10-20\%$, $20-30\%$ and $30-40\%$ centralities.  
 As stated in Ref~\cite{koch-error}, the error of the cumulants is intricate and significantly depended on the actual shape of the underlying multiplicity distribution.
 In addition, the error of $C_6/C_2$ is influenced by its measured value, as the same data sample is used to calculate both the measured
 $C_6/C_2$ and its error.  Consequently, the error-weighted average can lead to significantly different values of the cumulants.
 
\begin{figure}
\centering 
\includegraphics[width=4.6in]{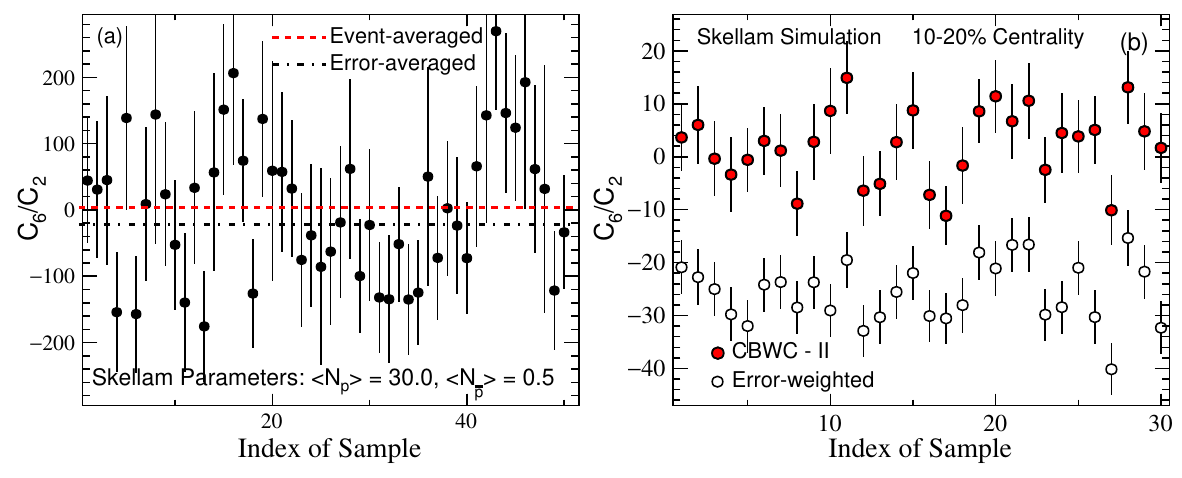}
\caption{\label{C62-error-10-20} (Color Online)  
(a): Skellam simulations of $C_6/C_2$  with the input parameters: 
$\left<N_{p}\right> = 30.0, \left<N_{\bar{p}}\right> = 0.5$. Each sample consisted of $10^5$ events.
The red and black dashed lines represent the event- and error-weighted averages over 50 samples, respectively.
(b): $C_6/C_2$ in 10-20\% centrality obtained by Skellam-based simulations of minimum bias data sample.  The data are identical to that used in Fig.~\ref{c62_c51_simulation}.
The results shown are for CBWC-II (red solid circles) and error-weighted (open circles) methods.
}
\end{figure}

 Simple Skellam simulations are employed to investigate this phenomenon. We initially randomly generate 50 independent samples,  each containing $10^{5}$ events following the 
 Skellam distribution with the input parameters:$\left<N_{p}\right> = 30.0, \left<N_{\bar{p}}\right> = 0.5$.  
 %As we know, despite having 100 million MB events at each energy, the statistical data within each Refmult3X bin remains relatively limited.
 %The simulated parameters and events  can be comparable to those observed at RHIC/BES-II.
 Results of the simulated $C_6/C_2$ are displayed in Fig.~\ref{C62-error-10-20}(a).
  In most scenarios,  the statistical error associated with $C_6/C_2$ tends
 to rise as the measured value of $C_6/C_2$ increases, as shown in Fig.~\ref{C62-error-10-20}(a).  
With the error-weighted factor being $\frac{1.0}{\left(\sigma_{C_6/C_2}\right)^2}$ as depicted in
 Eq.~(\ref{error_weight_error}), the error-weighted average of $C_6/C_2$ is  lower  that obtained by the event-weighted average.
  As a result, the  error-weighted average of $C_6/C_2$ is lower compared to the event-weighted average, illustrated by the red and black lines, respectively.
 
 %The red and black lines represent the event-weighted and error-weighted averages, respectively.
 
% It is evident that the error-weighted average of $C_6/C_2$ is 
 %significantly lower than . 
 %The measured $C_6/C_2$ in each $\delta 10\%$ centrality can not be calculated by error-weighted method.
 % notably lower than the event-weighted average.
 %This is because the error of $C_6/C_2$ is influenced by its measured value, as the same data sample is used to calculate both the measured
% $C_6/C_2$ and its error.

 In the central collisions of MB data,  there are numerous Refmult3x bins in every centrality, each containing a limited number of events.
 For instance, with 100M MB data in the UrQMD model at  $\sqrt{s_{NN}} = 11.5$ GeV, 
 there are already over 100 Refmult3X bins in the 10-20\% centrality, with most of them containing fewer than 0.1M events.
 Consequently, the effects of the error-weighted in the MB data sample in 10-20\% centrality may similar than that shown in Fig.~\ref{C62-error-10-20}(a).
 Using the Skellam-based simulation data  identical to that in  Fig.~\ref{c62_c51_simulation}, 
  Fig.~\ref{C62-error-10-20}(b) illustrates the difference of  $C_6/C_2$  in 10-20\% centrality between  event- and error-weighted averages.
  It reveals that the measured $C_6/C_2$ obtained via event-weighted averages exhibit random fluctuations from the Skellam expectation of unity. 
  Conversely, the values of $C_6/C_2$ calculated using error-weighted averages are significantly lower compared to those obtained through event-weighted averages and the Skellam expectation.
Therefore, with the RHIC/BES-II statistics, the error-weighted average method  cannot be applied to calculate  $C_6/C_2$ in each $\delta 10\%$ centrality in central collisions.  

If  the events-weighted method is initially utilized to calculate $C_6/C_2$ ($C_5/C_1$) in each $\delta 10\%$ centralities.
Subsequently, their values in the 0-40\% centrality are weighted according to their respective errors, leading to the development of the proposed CBWC-III method.
 Although the error-weighted
averages can lead to much smaller uncertainties,  it does not imply that the CBWC-III method is a superior option for computing 
 $C_6/C_2$ and $C_5/C_1$ in $0-40\%$ centrality. 
 This is due to the characteristics of their statistical uncertainties.
 In a normal distribution, the delta theorem method shows that the error of $C_6/C_2$ is : 
 \begin{equation}\label{error-C6C2-C5}
error\left(C_6/C_2\right) \propto \frac{\sigma^4}{\sqrt{n}}
\end{equation}
Here $\sigma$ is the standard deviation of the net-proton multiplicity distributions. 
From peripheral to central collisions, there is a noticeable increase in the standard deviation,  as evident in Fig.~\ref{c62_cbwc_urqmd}.
The minimum bias data sample shows nearly identical numbers of events in the
 0-10\%,  10-20\%, 20-30\% and 30-40\% centralities. Therefore, there is a significant rise in the associated statistical error from peripheral to central collisions.
   When employing the error-weighted scheme illustrated in Eq.~(\ref{error_weight_error}) to determine the results of hyper-order cumulants
  in 0-40\% centrality,  the results in the 0-10\% centrality  can be almost negligible.

\begin{figure}
\centering
\includegraphics[width=4.3in]{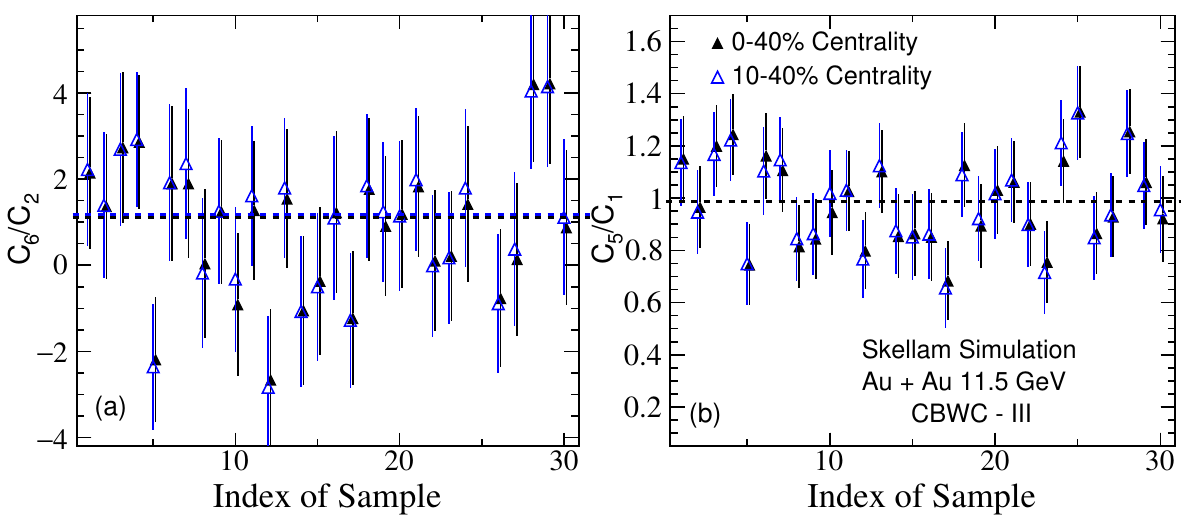}
\caption{\label{C62-UrQMD-simulations} (Color Online) 
Skellam-based simulations  results of net-proton $C_6/C_2$ (left panel) and $C_5/C_1$ (right panel) calculating by using CBWC-III  method.
The results shown are for $0-40\%$ (solid circles)  and $10-40\%$ (open circles) centralities.  }
\end{figure}

By using CBWC-III method, Fig.~\ref{C62-UrQMD-simulations}  shows the Skellam-based simulation results of net-proton $C_6/C_2$ (a) and $C_5/C_1$ (b) in 0-40\% and 10-40\% Au + Au collisions centralities
 at $\sqrt{s_{NN}}$ = 11.5 GeV.
% The results corresponding to indices 1 to 30 are derived from Skellam simulations, utilizing the same data as in Fig.~\ref{c62_c51_simulation}. The results for index 32 are obtained from the UrQMD model.
 As anticipated,  
 the statistical uncertainties of $C_6/C_2$ and $C_5/C_1$ in 0 - 40\% centrality 
obtained by the CBWC-III method are significantly smaller for each sample, comparing to those obtained by CBWC-II method shown in 
 Fig.~\ref{c62_c51_simulation}. 
 However, 
the values of both $C_6/C_2$ and $C_5/C_1$ in 0-40\% and 10-40\% central collisions are almost the same, 
with the central value measured and the statistical uncertainty associated being indistinguishable.
The measurement of the cumulant ratio in $0-10\%$ centrality is widely recognized as crucial for studying the QCD phase transition, 
 including $C_4/C_2$ of net-proton number, net-charge and net-kaon,  among others~\cite{star-proton, star-charge, star-kaon}.
 Therefore, 
 when employing the CBWC-III method with the MB trigger data sample, 
 it is essential to carefully assess   the measured $C_6/C_2$ and $C_5/C_1$.

\subsection{Comparisons  in the UrQMD model}  
We start by summarizing the factors considered to evaluate the method's suitability in our research.
Initially, the result should align  with theoretical expectations and not be underestimated. 
Additionally, as it is most probable to reach thermal equilibrium and undergo a phase transition in the most central collisions (0-10\%), 
 the result in 0-10\% centrality hold the most significance and should not be disregarded.
 Lastly, combined with these two conditions, having smaller statistical errors is preferred, particularly due to the limited statistics available for analyzing these two ratios.

\begin{figure}
\centering
\includegraphics[width=4.3in]{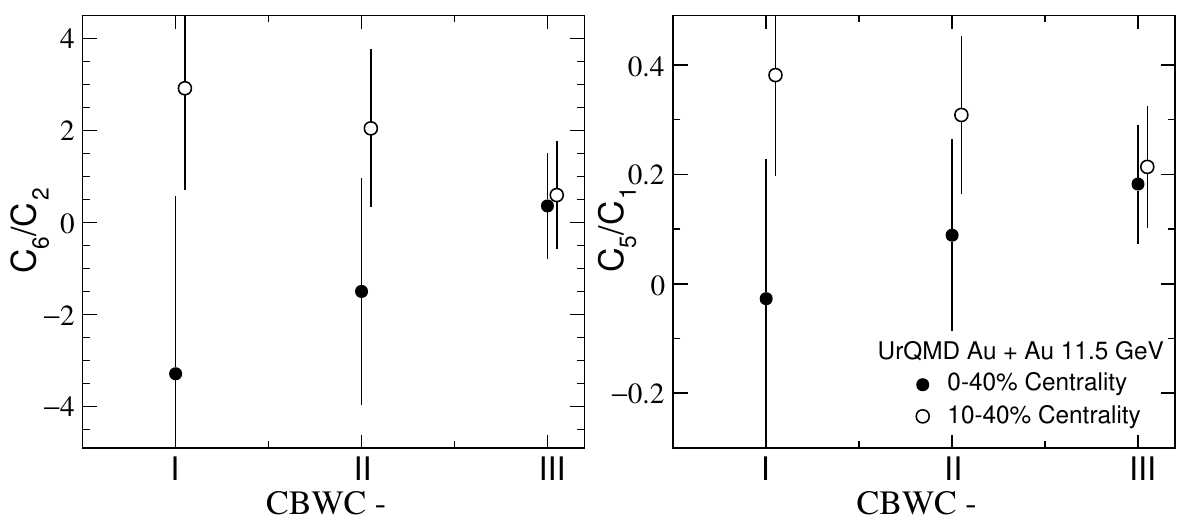}
\caption{\label{C62-UrQMD} (Color Online)  UrQMD results of net-proton $C_6/C_2$ (left panel) and $C_5/C_1$ (right panel) calculating by using CBWC-I, -II, and -III, respectively.
The results shown are for $0-40\%$ (solid circles)  and $10-40\%$ (open circles) centralities.  
  }
\end{figure}

 Fig.~\ref{C62-UrQMD} shows the net-proton $C_6/C_2$ and $C_5/C_1$ in 
 0-40\% and 10-40\% centralities
 at $\sqrt{s_{NN}}$ = 11.5 GeV in the UrQMD model  using the CBWC-I, -II, and -III methods. 
 With the CBWC-III method, the results of these two ratios in 0-40\% and 10-40\% centralities are found to be identical. It means that the result in 0-10\% centrality is disregarded. 
 On the contrary, the values of these two ratios in the 0-40\% centrality are notably lower than those in the 10-40\% centrality. This discrepancy is attributed to the significantly lower results of these ratios in the 0-10\% centrality compared to the 10-20\%, 20-30\%, and 30-40\% centralities in the UrQMD model, as illustrated in Fig.~\ref{c62_cbwc_urqmd}. 
 This indicates that the results in the 0-10\% centrality play a significant role in determining the final measured results in the 0-40\% centrality.

 Fig.~\ref{C62-UrQMD} also
  illustrates that the errors of these two ratios calculated using CBWC-II are notably smaller than those calculated using CBWC-I.
  Furthermore, as demonstrated in Fig.~\ref{c62_c51_simulation}, the values of these ratios align with the theoretical expectations when computed using both methods. 
  As a result, the CBWC-II method is deemed more appropriate for data analysis.

 Of course, here we propose using the CBWC-II method to calculate these two ratios within a large 0-40\% centrality when employing the CBWC method.
 However,  the UrQMD results in Fig.~\ref{c62_cbwc_urqmd} show that  these two ratios in 0-10\% centrality
 differ significantly from those in other centralities. 
 Consequently, fundamentally, the results obtained by a higher statistics sample of events in 0-10\% centrality are  still remain  the most valuable  for studying the QCD phase transition. 
 
\section{Summary}

The hyperorder cumulants are proposed to study the QCD phase transition. 
To minimize uncertainty, experimental analysis focuses on studying the energy dependence within a broader centrality bin width of 0-40\%, rather than narrower widths like 0-10\% or 0-5\%. 
This expanded centrality bin width makes the hyperorder cumulants more sensitive to the schemes of the CBWC method, 
which are utilized to suppress initial volume fluctuations.
Variations in volume cancellation methods or weighting schemes within the CBWC formula can result in notable discrepancies in the obtained results.

We examine the impact of various CBWC methods on the net-proton $C_{5}/C_{1}$ and $C_6/C_2$ 
measurements in Au + Au collisions using the UrQMD model and Skellam-based simulations. Utilizing MB data samples to calculate these ratios in the 0-40\% centrality, it 
 reveals that the error-weighted average
  can not be used to calculate these two cumulant ratios in any scenario. 
  Directly applying the error-weighted average in
   the 0-10\%, 10-20\%, 20-30\%, and 30-40\% centralities can result in underestimated measurements. 
    In contrast, the event-weighted method is initially employed to calculate 
 $C_6/C_2$ ($C_5/C_1$) in those four centralities.
Subsequently, the $C_6/C_2$ ($C_5/C_1$) values in the 0-40\% centrality are weighted according to their respective errors (CBWC-III). 
   This approach can lead to the
 $C_{5}/C_{1}$ and $C_6/C_2$ values in the 0-10\% centrality being almost negligible. 
 The results in the 10-40\% and 0-40\% centralities are indistinguishable, encompassing both the measured central value and its associated statistical uncertainty. 
 This lack of distinction is not conducive for using these two ratios to analyze the QCD phase transition, 
 as it is more probable to approach thermal equilibrium and undergo a phase transition in the most central collisions.

By employing the event-weighted scheme, the direct calculation of $C_m/C_n$ 
in each multiplicity bin (CBWC-II) is favored over the separate computation of $C_m$ and $C_n$
in each bin (CBWC-I). 
The formula of the CBWC-II method aligns more closely with the theoretical  rationale of eliminating volume contributions.
In our research, for eight centralities  from 0-10\% to 70-80\%,  the discrepancies in the measured values of $C_6/C_2$ (and $C_5/C_1$)
between these two methods are negligible.
However, with a wider centrality bin width of 0-40\%, the errors for both $C_5/C_1$ and $C_6/C_2$ obtained using the CBWC-II method 
are significantly reduced, approximately  60\% $\sim$ 70\% compared to those obtained through CBWC-I method.
This impact of the decreased uncertainty  is equivalent to doubling the statistical sample in experiment. 

%There is a noticeable pattern in the variation between $\left(C_6/C_2\right)_{(C-I)}$ and $\left(C_6/C_2\right)_{(C-II)}$, 
%(and also $\left(C_5/C_1\right)_{(C-I)}$ and $\left(C_5/C_1\right)_{(C-II)}$).
 %Near the average value, the difference between them is small.
%Above the average value, the majority of the measured  $\left(C_6/C_2\right)_{(C-II)}$ values consistently
 %lie below their corresponding $\left(C_6/C_2\right)_{(C-I)}$.
% Conversely, for values below the average, the majority of the measured  $\left(C_6/C_2\right)_{(C-II)}$ 
 %values are consistently higher than $\left(C_6/C_2\right)_{(C-I)}$.  
 %Specifically,  outside $1\sigma$ range of the average, 
  %the measured $\left(C_6/C_2\right)_{(C-II)}$ and $\left(C_6/C_2\right)_{(C-I)}$ are significantly different.
  %Experimentally, 
  %it can serve as supplementary validation to determine if the measured value closely matches, falls below, 
  %or exceeds the expected values in the experiment. This can aid in refining the range of expected values without requiring additional statistics.

The statistics in our calculations can be compared with RHIC/BES-II data. 
The centrality has already been determined using Refmult3X. 
Therefore, our results and discussions provide a straightforward reference for the measurements of $C_5/C_1$ and $C_6/C_2$ in relativistic heavy-ion collisions.  

\section{Acknowledgements}
This work is supported by the National Key Research and Development Program of China (No. 2022YFA1604900); the National Natural Science Foundation of China (No. 12275102), 
and the Scientific Research and Innovation Team Program of Sichuan University of Science and Engineering (No. SUSE652A001).
We also acknowledge the High Performance Computing Center of Nanjing University of Information Science and Technology for their support of this work.

\clearpage
%\end{CJK*}
\end{document}